\documentclass[%
 reprint,
 amsmath,amssymb
 aps,
]{revtex4-1}

\usepackage{graphicx}% Include figure files
\usepackage{dcolumn}% Align table columns on decimal point
\usepackage{bm}% bold math
\usepackage{color}
\usepackage{wasysym}
\usepackage{epstopdf}
\usepackage{verbatim}
\usepackage{natbib}
\usepackage{epsfig}
\usepackage{subfigure}
\begin{document}

\preprint{APS/123-QED}

\title{Quantum-Enhanced Velocimetry with Doppler-Broadened Atomic Vapor}% Force line breaks with \\

\author{Zilong Chen$^1$}%
 \email{chen$_$zilong@ntu.edu.sg }
\author{Hong Ming Lim$^1$, Chang Huang$^1$, Rainer Dumke$^{1,2}$}
\author{Shau-Yu Lan$^1$}%
 \email{sylan@ntu.edu.sg}
 \affiliation{%
$^{1}$Division of Physics and Applied Physics, School of Physical and Mathematical Sciences, Nanyang Technological University, Singapore 637371, Singapore\\
$^{2}$Centre for Quantum Technologies, National University of Singapore, 3 Science Drive 2, Singapore 117543, Singapore}


%\affiliation{
%\affiliation{
 %Third institution, the second for Charlie Author
%\affiliation{%


\date{\today}% It is always \today, today,

\begin{abstract}
Traditionally, measuring the center-of-mass (c.m.) velocity of an atomic ensemble relies on measuring the Doppler shift of the absorption spectrum of single atoms in the ensemble. Mapping out the velocity distribution of the ensemble is indispensable when determining the c.m. velocity using this technique. As a result, highly sensitive measurements require preparation of an ensemble with a narrow Doppler width. Here, we use a dispersive measurement of light passing through a moving room temperature atomic vapor cell to determine the velocity of the cell in a single shot with a short-term sensitivity of 5.5 $\mu$m s$^{-1}$ Hz$^{-1/2}$. The dispersion of the medium is enhanced by creating quantum interference through an auxiliary transition for the probe light under electromagnetically induced transparency condition. In contrast to measurement of single atoms, this method is based on the collective motion of atoms and can sense the c.m. velocity of an ensemble without knowing its velocity distribution. Our results improve the previous measurements by 3 orders of magnitude and can be used to design a compact motional sensor based on thermal atoms.

\end{abstract}

\pacs{Valid PACS appear here}% PACS, the Physics and Astronomy
\maketitle
Measuring motion of atoms plays a significant role in performing high precision inertial sensing, such as gravity, gravity gradient, and rotation \cite{Cro}.  It has also been used to study fundamental physics, including quantum tests of the equivalence principle \cite{Sch,Zho}, and measurements of the fine structure constant \cite{Par} and Newton’s constant $G$ \cite{Ros}. Current atoms-based motional sensors rely on measuring the first-order Doppler shift of the absorption spectrum of some narrow linewidth transition of single atoms in a large thermal ensemble. One method is the Doppler sensitive two-photon Raman velocimetry that uses a pair of counterpropagating laser fields to drive a pair of long-lived states of atoms \cite{Kas}. By detuning the relative frequency of the counterpropagating laser fields, a subgroup of atoms with finite velocity width, which is determined by the duration of the pulse length, can be selected. Because of the finite temperature of the ensemble, the c.m. velocity is then determined by scanning the detuning of the laser fields to map out the Doppler distribution and fit the one-dimensional the Maxwell-Boltzmann distribution with the data as shown in Fig. 1 (left). The sensitivity is, therefore, largely limited by the Doppler broadening of the atomic ensemble used. To improve the sensitivity, one would have to prepare an ensemble at ultralow temperature \cite{Kov}, which requires a complex laser cooling and trapping setup.

\begin{figure}[h]
\subfigure{\label{fig:1a}\includegraphics[scale=0.18]{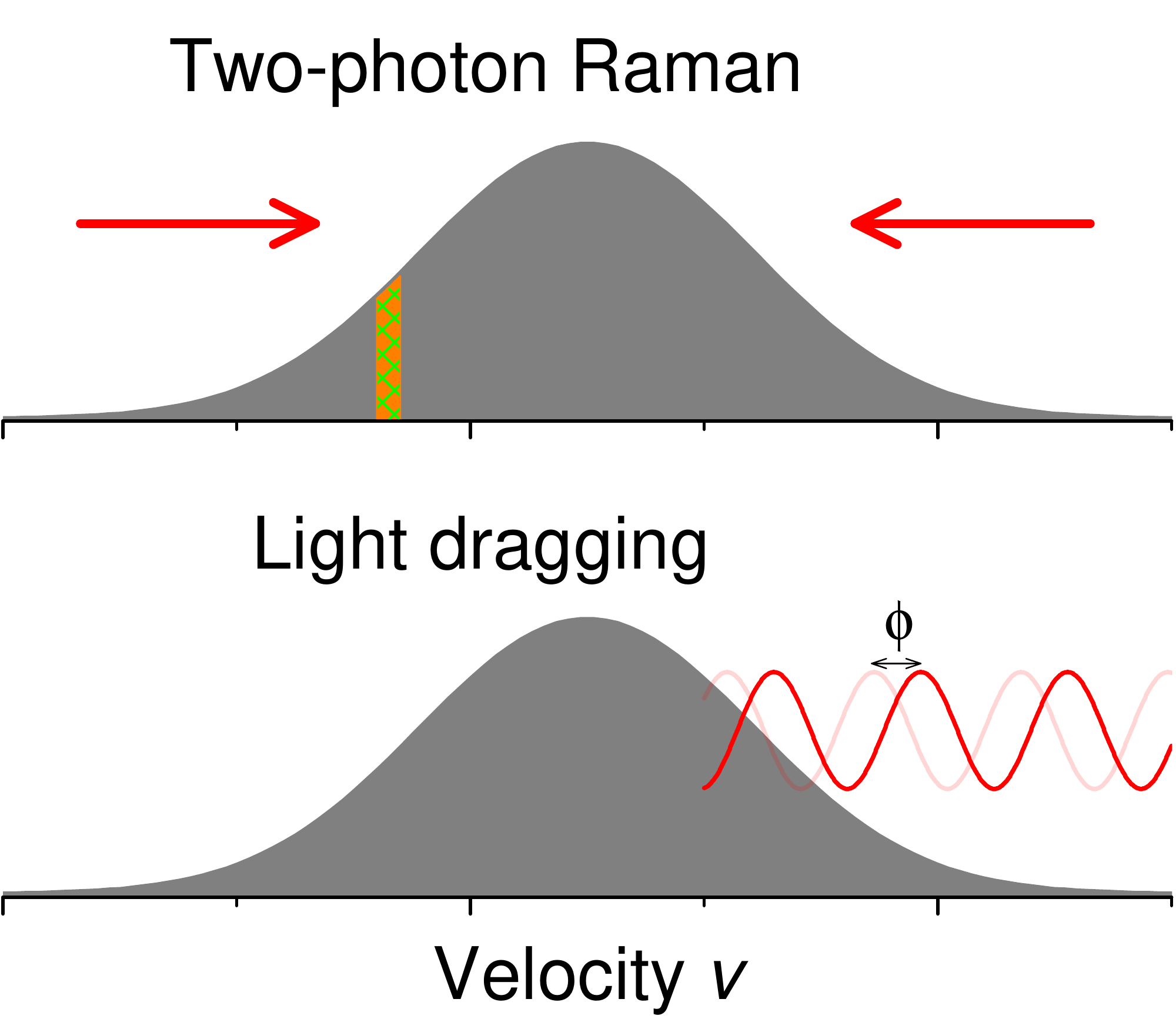}}
\subfigure{\label{fig:1b}\includegraphics[scale=0.15]{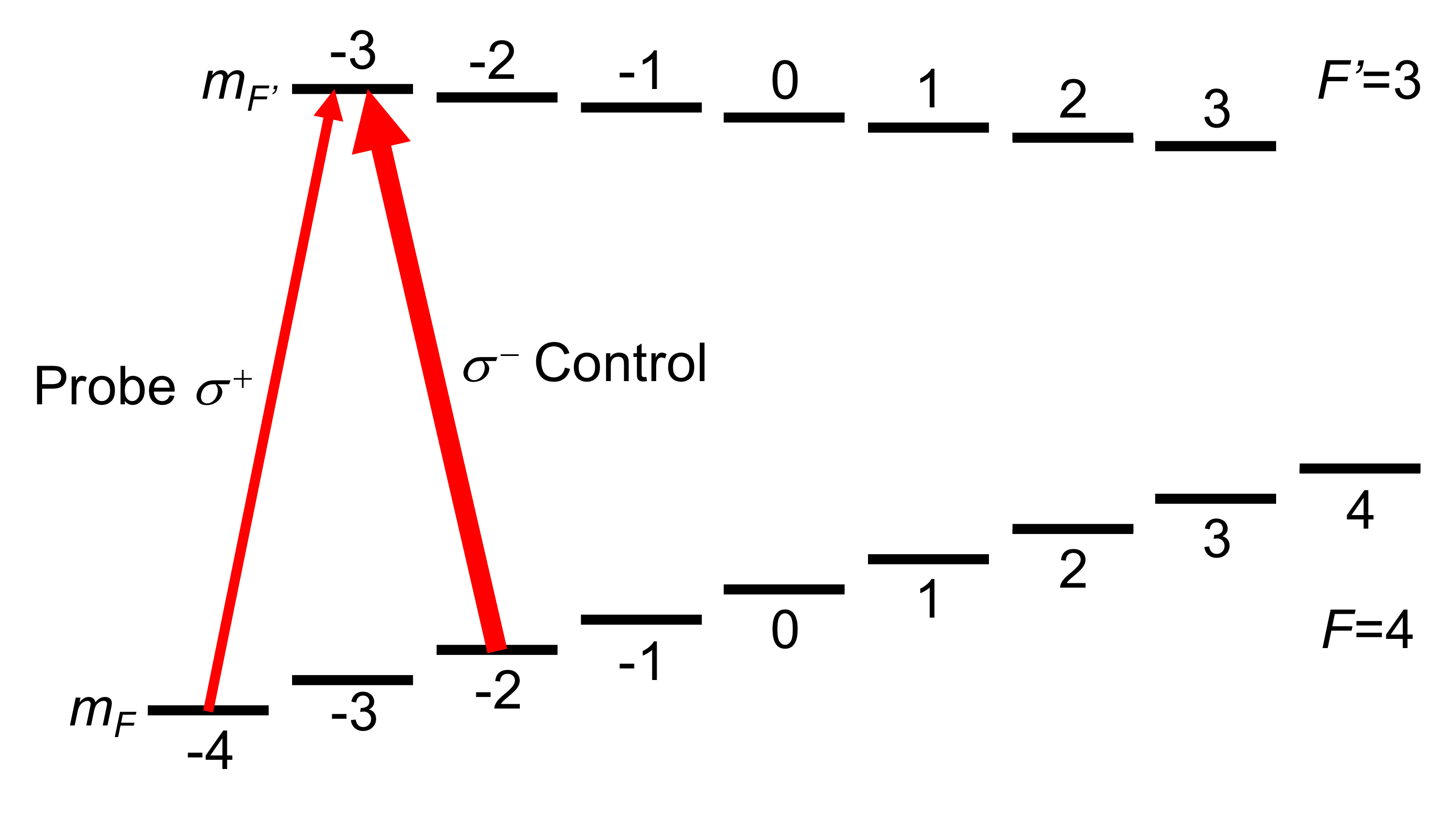}}
\caption{Concept illustration and EIT energy levels. (Left) Illustration of the comparison of two-photon Raman velocimetry in cold atoms and the light-dragging velocimetry. The shaded area is the population of one-dimensional Maxwell-Boltzmann distribution of an atomic ensemble. The orange mesh area indicates the velocity width that each Raman pulse selects. The light-dragging method measures the phase shift $\phi$ of transmitted light to determine the c.m. velocity. (Right) Energy levels of Cs D1 line addressed by the control and probe fields at 895 nm for EIT. The Zeeman energy levels are lifted by an external magnetic field. The control and probe fields are generated by a single diode laser and split by a polarization beam splitter. Thicker red arrow indicates stronger optical power.}
\label{fig:1}
\end{figure}

Warm atomic vapor cells have been applied in optical magnetometers \cite{Bud}, atomic clocks \cite{Cam}, and inertial sensing \cite{Bie}. The compact and versatile features of the apparatus make them excellent candidates for deployable high precision sensing devices. While most of the applications utilize stationary vapor cells, the recent demonstration of measuring the motion of a moving atomic vapor cell displays a way of applying an atomic vapor cell as a motional sensor \cite{Saf}. Here, we exploit quantum interference in electromagnetically induced transparency (EIT) effect for a moving Doppler-broadened atomic vapor cell and determine the velocity of the vapor cell by measuring the phase shift of light passing through the moving medium \cite{Saf,Zim,Sha,Dav}. We demonstrate the sensitivity of the velocity of the atomic vapor to the level of 31 $\mu$m s$^{-1}$ after 32 ms of integration time or equivalently 5.5 $\mu$m s$^{-1}$ Hz$^{-1/2}$ for short integration times.

When light propagates in a moving medium, its phase velocity, to the first order of Lorentz transformation \cite{Saf}, can be written as
\begin{equation}
v_{\textrm{p}}=\textrm{c}/n\pm F_{\textrm{d}}v,
\end{equation}
where $\textit{F}_{\textrm{d}}$=1$-$1/$\textit{n}^{2}$+($\omega/n^{2}$)[$\partial n$($\omega$)/$\partial\omega$] is the dragging coefficient, $\textrm{c}$ is the speed of light in vacuum, $n$ is the index of refraction of the medium, $v$ is the speed of the medium, and $\omega$ is the angular frequency of the light in the laboratory frame. This dependence was called the light-dragging effect by H. Fizeau two centuries ago \cite{Fiz} and later explained as the first-order Doppler shift by special relativity \cite{Lor}. Although $n$ is typically close to unity, this effect can be enhanced in a highly dispersive medium \cite{Lau}. The dragging coefficient can be simplified by the group index $n_{\textrm{g}}\equiv$$\textrm{c}/v_{\textrm{g}}$ as $F_{\textrm{d}}$=1$-$1/$n^{2}$+($n_{\textrm{g}}-n$)/$n^{2}$, where $v_{\textrm{g}}$=$\textrm{c}$/($n$+$\omega$($\partial n$($\omega$)/$\partial\omega$)) is the group velocity. For $n\simeq$1 and large group index, $F_{\textrm{d}}$ is simply the ratio of the speed of light and group velocity. In general, when light excites an atomic transition resonantly, the dispersive property of atom-light interaction modifies the index of refraction and also causes a strong absorption of light by the medium. Various methods of detecting motion and rotation using a dispersive medium have been proposed \cite{Zim,Sha,Dav}. Recently, measurements of the phase shift of light through a moving medium have been demonstrated in a hot Rb vapor cell by sending a single laser far detuned from the atomic resonance to avoid large absorption \cite{Saf}. A moving cold atomic ensemble under the EIT condition is further used to increase the sensitivity while minimizing the absorption compared to the hot vapor experiment \cite{Kua}. We also note that measurements of group velocity dragging under EIT have been shown in a hot vapor by selecting a velocity group of atoms through optical pumping \cite{Str}.

When light passes through a medium of length $L$, it accumulates a phase of $\Phi$=$\textit{k}'L$=$L\omega$/$v_{\textrm{p}}$=$L\omega$/($\textrm{c}$(1+$v$/$v_{\textrm{g}}$))$\simeq$($L\omega$/$\textrm{c}$)(1$-v$/$v_{\textrm{g}}$) when $v\ll v_{\textrm{g}}$ , where $k'$ is the wave number of light in the atoms' frame.  The velocity of the medium can then be determined by measuring the phase shift of the light relative to a reference field that does not pass through the medium as
\begin{equation}
\phi=-kLv/v_{\textrm{g}}=-k\tau v,
\end{equation}
where $k$=2$\pi$/$\lambda$ is the wave number of the light in the lab frame, and $\tau$ is the group delay time. The large group delay of slow light under the EIT condition can boost the phase shift and lead to a highly sensitive measurement of the velocity.

Our $\Lambda$ type EIT configuration is formed by the Zeeman states of the Cs hyperfine ground state 6S$_{\textrm{1/2}}$, $F$=4, and the excited state 6P$_{\textrm{1/2}}$, $F'$=3 as shown in Fig. 1 (right). The experimental apparatus is based on two external cavity diode lasers (ECDLs), and a Cs vapor cell maintained at $25.4^\circ$C. One of the ECDLs is locked to 80 MHz red detuned of $F$=4 to $F'$=3 and split into two different paths for the probe and the control fields, and the other one is locked to $F$=3 to $F'$=4 for optical pumping to the $F$=4 state. The control field is diffracted by an acoustic-optical modulator (AOM) at 79 MHz modulation frequency before being coupled into a single mode fiber. The probe field is diffracted by another AOM at 80 MHz modulation frequency. The zeroth-order output of 1.5 $\mu$W is for heterodyne detection, and the +1-order output of 70 nW is for the probe field. They are recombined and coupled into another single mode fiber to form an 80 MHz beat note, as shown in Fig. 2 (top). In order to induce a two-photon phase shift in a copropagating EIT configuration, the output of the control field fiber is mounted on the motorized translation stage of the vapor cell while the output of the probe field fiber is placed on the optical table. Half of the probe field power is split by a half-wave plate and polarization beam splitter (PBS) to serve as a reference signal. The probe and control fields are then combined on a PBS followed by a quarter-wave plate (QWP) on the translation stage to adjust the polarization of the probe and control fields to $\sigma^{+}$ and $\sigma^{-}$ polarization as shown in Fig. 2 (bottom).

\begin{figure}[h]
\subfigure{\label{fig:2a}\includegraphics[scale=0.3]{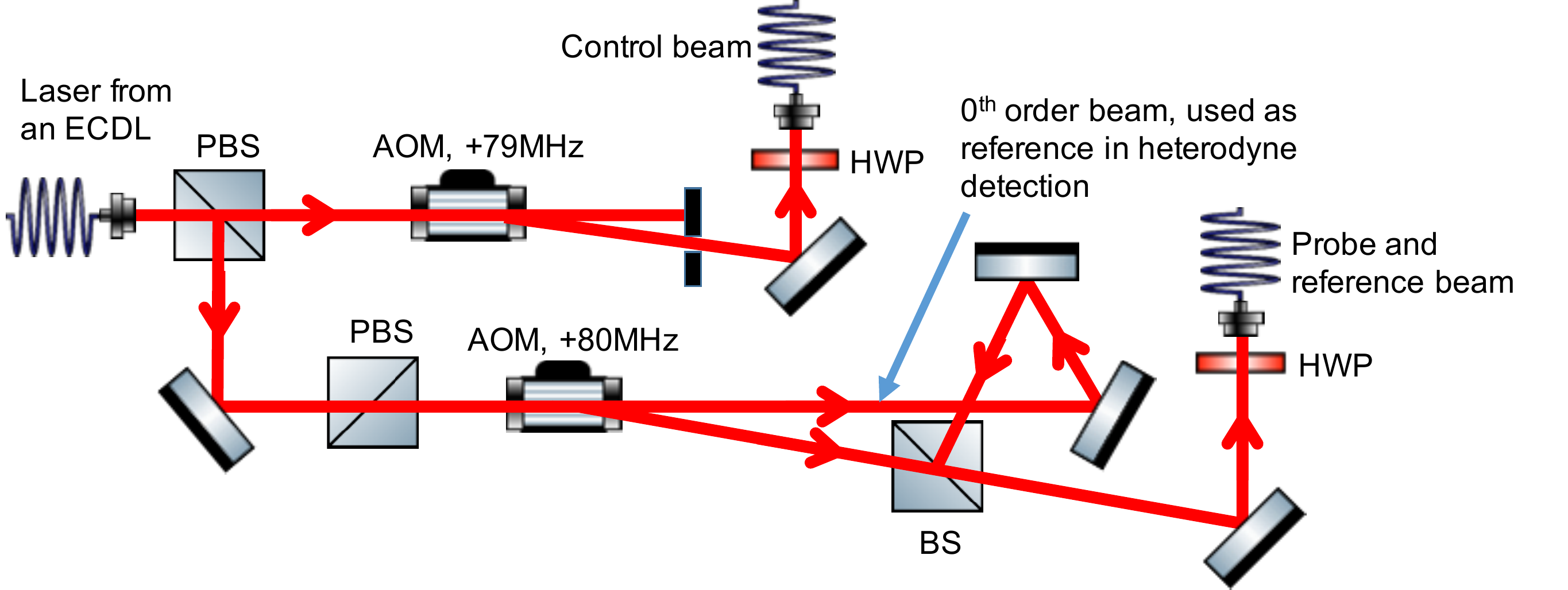}}
\subfigure{\label{fig:2b}\includegraphics[scale=0.3]{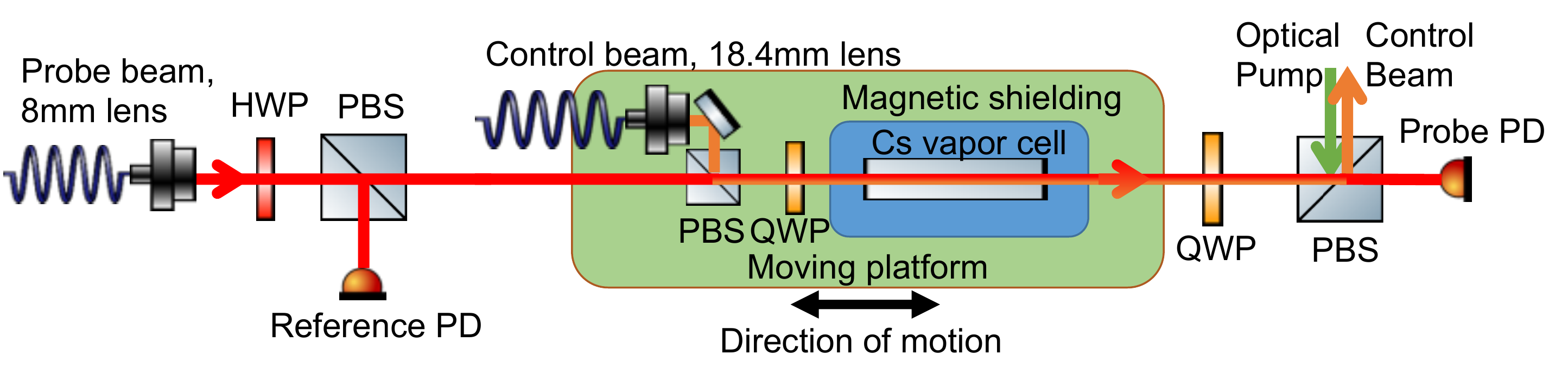}}
\caption{Experimental configuration. (Top) Schematics of frequency generation of the probe, reference, and control fields for phase shift measurements. HWP: half-wave plate. BS: beam splitter. (Bottom) Experimental setup. The green rectangle is a moving optical breadboard connected to a stepper motor, and the blue rectangle represents the one layer G-Iron metal enclosure to shield the stray magnetic field. The control beam fiber coupler and the vapor cell are fixed on the breadboard to induce the two-photon detuning when the platform is moving. PD: photodiode.}
\label{fig:2}
\end{figure}

The strong control field of 220 $\mu$W also serves as an optical pumping beam to pump atoms to the $m_{\textrm{F}}$=$-$4 and $-$3 state. The 1/$e^{2}$ beam diameters of the control and probe fields are approximately 4.0 mm (collimated by a lens with 18.4 mm focal length) and 1.8 mm (collimated by a lens with 8 mm focal length), respectively. After passing through the cell, another pair of QWP and PBS is used to separate the probe field from the control field, and the probe field is then detected by an avalanche photodetector with a 3 dB bandwidth of 50 MHz. The translation stage is driven by a stepper motor with a resolution of 2000 steps per revolution and a linear worm gear with 5 mm translation per revolution. Using an Arduino Uno to control the stepper motor driver, we verify the translation stage velocity to be $v$=5 mm$\times$motor step rate/2000. The cell is mildly magnetic shielded to minimize the influence of the ambient magnetic field by a factor of about 25. A solenoid is placed inside the shield to supply a homogeneous magnetic field in the axial direction to break the Zeeman state degeneracy and also adjust the two-photon resonance condition. The energy shift between adjacent Zeeman states is about 1 MHz.

Figure 3 shows the absorption spectrum of the probe field under EIT condition when the vapor cell is stationary. The two-photon detuning between the three-level system, and the control and probe fields is scanned by the external magnetic field. The data without optical pumping arefitted with a Lorentzian function, and the full width at half maximum (FWHM) is 66.9(3) kHz. The data with optical pumping are fitted better with a Gaussian function, and the FWHM is 58.8(5) kHz. Considering the root-mean-square (rms) speed $v_{\textrm{rms}}$=$\sqrt{k_{\textrm{B}}T/m}$=136 m s$^{-1}$ of atoms at $25.4^\circ$C, where $k_{\textrm{B}}$ is the Boltzmann constant and $m$ is the Cs mass, the transit time of atoms through the probe beam area is about 13 $\mu$s, which corresponds to an EIT linewidth of 1/$\pi$/(1.3$\times$10$^{-5}$)=24.4 kHz. The discrepancy from our measured EIT linewidth is attributed to the ground state's decoherence from the residual magnetic field inhomogeneity and power broadening from the control field.

\begin{figure}[h]
 \includegraphics[scale=0.3]{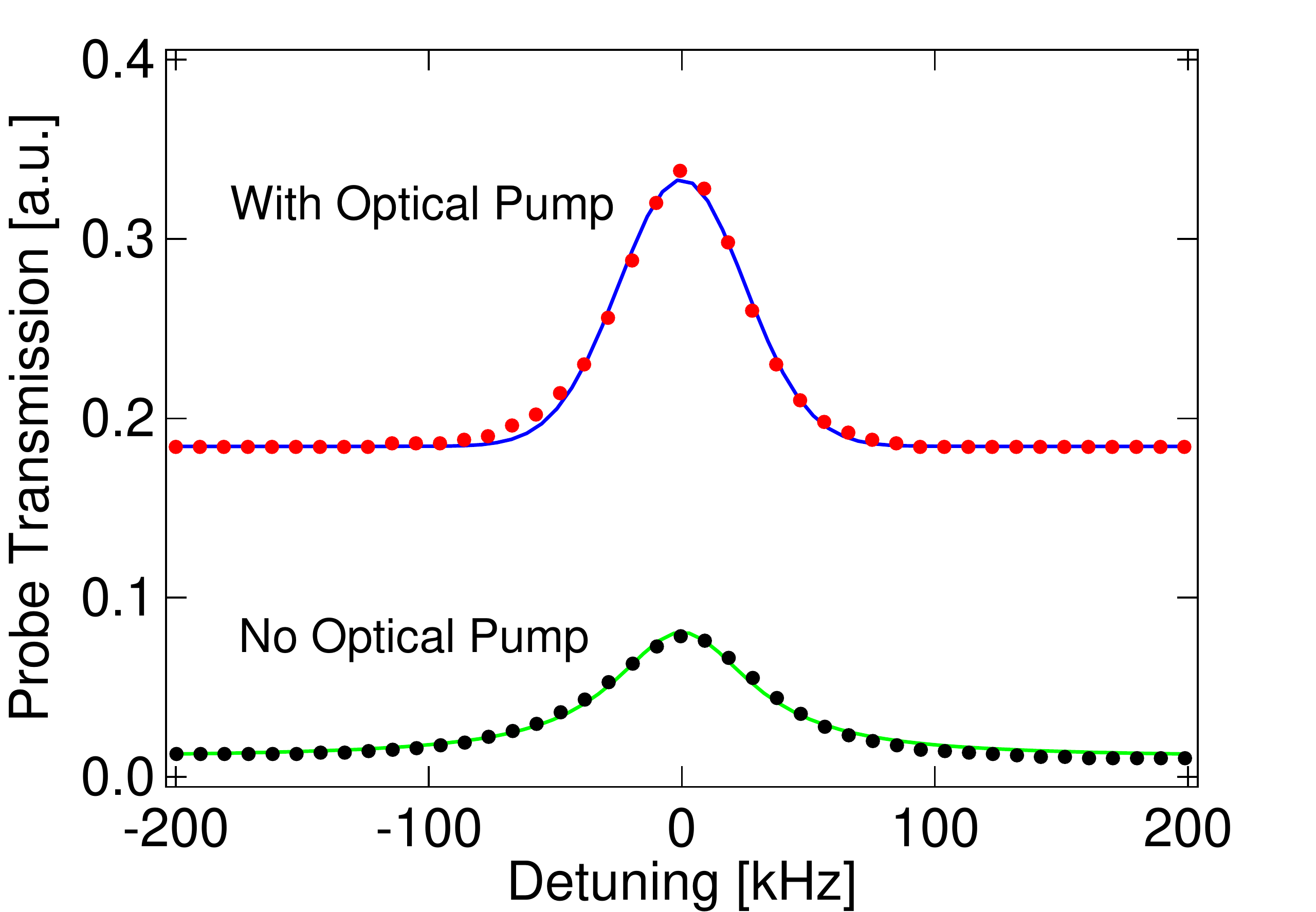}
  \caption{EIT linewidth measurement. Transmission of probe field as a function of two-photon detuning of the three-level EIT system by scanning the external magnetic field. A Gaussian fit (blue curve) to the probe transmission data with optical pump (red points) yielding a full width at half maximum of 58.8(5) kHz. A Lorentzian fit (green curve) to the probe transmission data without optical pump (black points) yielding a full width at half maximum of 66.9(3) kHz. For display purposes, only every 20th data point is plotted. Each data point is the average of 16 scope traces.}
  \label{fig:3}
\end{figure}

\begin{figure}[h]
 \subfigure{\label{fig:4a}\includegraphics[scale=0.3]{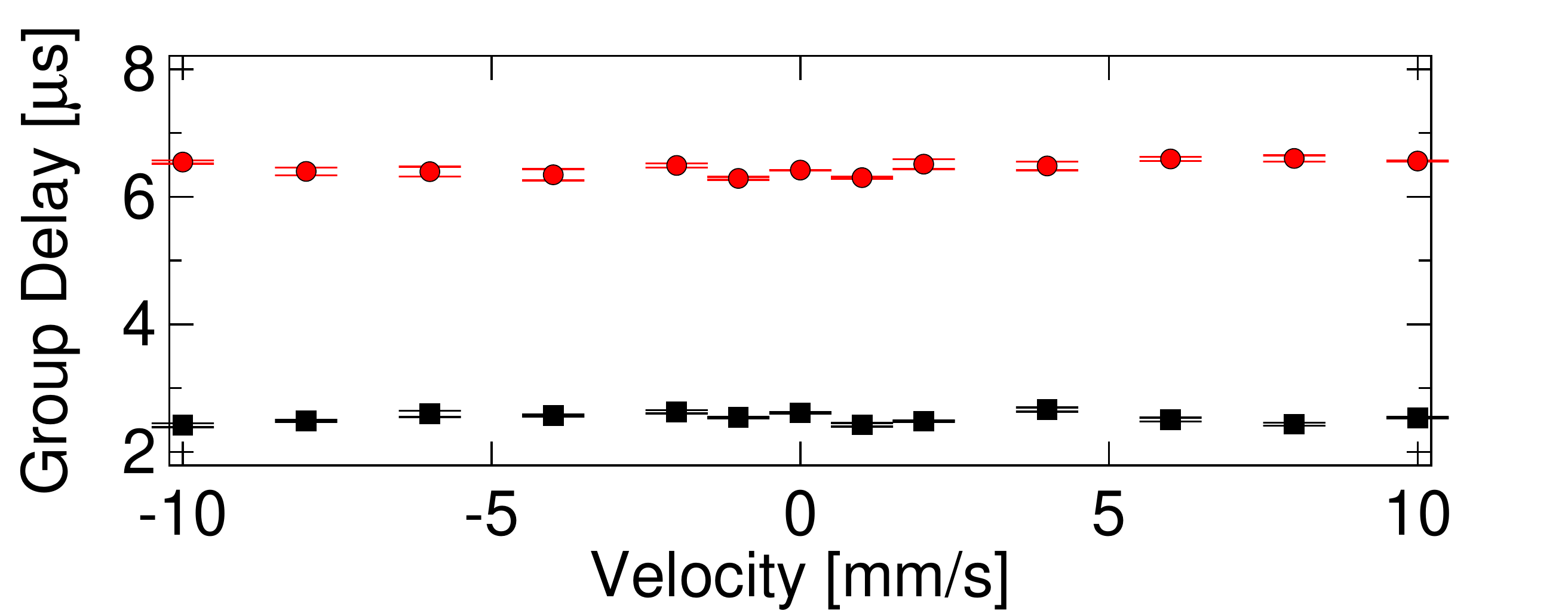}}
 \subfigure{\label{fig:4b}\includegraphics[scale=0.3]{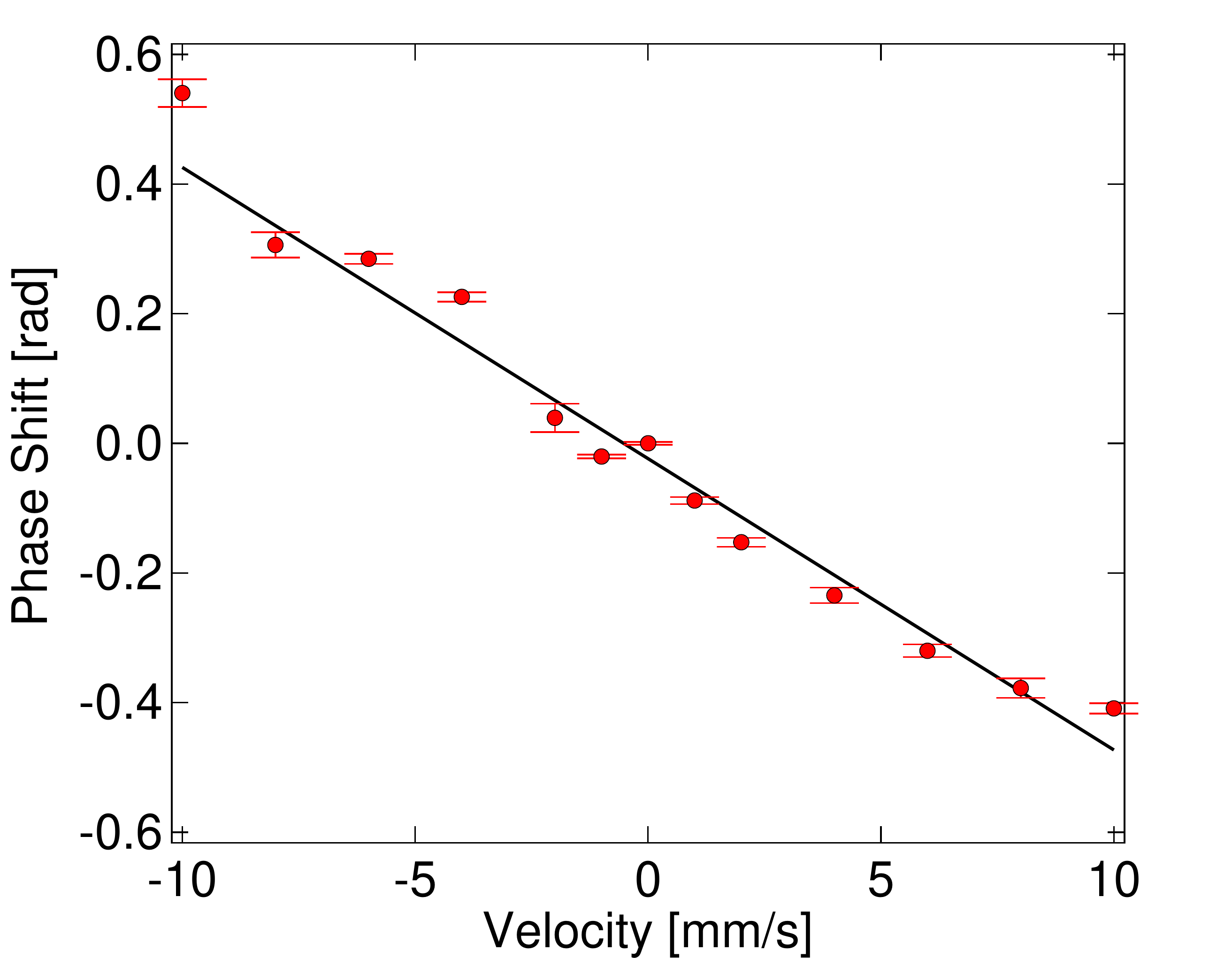}}
\caption{Phase shift measurement. (Top) Group delay measurements as a function of the velocity of the moving vapor cell under the EIT condition with optical pump (red circles) and without optical pump (black squares). Each data point is the average of five experimental runs. (Bottom) Phase shift measurements as a function of the velocity of the moving vapor cell under the EIT condition and with optical pump (red points). A linear fit (black line) to the data yields a fitted slope of $-$44.9(5) mrad per mm s$^{-1}$, in agreement with the theoretical value of $-$45.3(3) mrad per mm s$^{-1}$ calculated from the average group delay of 6.46(4) $\mu$s over the velocity range used here. The velocity sensitivity near zero velocity is estimated to be 50 $\mu$m s$^{-1}$ from the standard error of five measured phase shift fluctuations at zero velocity. Each data point is the average of five experimental runs.}
\label{fig:4}
\end{figure}

The velocity of the vapor cell is determined by measurements of group delay and phase shift of the probe field via Eq. (2) where $k$ is taken as 2$\pi$/895 nm$^{-1}$. To measure the group delay, the probe signal intensity is modulated at 10 kHz and averaged 16 times with a 1 ms span on the oscilloscope. We measure the phase shift between the probe field under the EIT condition and without the EIT condition (by blocking the control beam) by fitting to a sine function where the frequency is fixed to 10 kHz. The group delay is then determined from the measured phase shift and the 10 kHz modulation frequency. Figure 4 (top) shows that the group delay at different vapor cell velocities agrees within a few percent. The cell temperature is maintained at $25.4^\circ$C. With the length of the vapor cell $L$=7 cm and 6.46(4) $\mu$s of group delay, the dragging coefficient $F_{\textrm{d}}$ in our experiment is about 2.8$\times$10$^{4}$. The phase measurements are done by comparing the phase of the probe and reference fields at each velocity.  To avoid the influence of imperfect extinction of the control field by the final PBS on the probe detector, the 80 MHz probe beat note is demodulated to 50 kHz by mixing the beat note with a 79.95 MHz signal and averaged 16 times on the oscilloscope with 100 $\mu$s time span. The parasitic 79 MHz reference beat note that arises from the control leakage is instead demodulated to 950 kHz and strongly attenuated with a 100 kHz single-pole low pass filter. We compare the results with the motor velocity setting, as shown in Fig. 4 (bottom). The $x$ axis indicates 13 different settings of the motor velocity, and the $y$ axis shows the values of our phase measurements. The black line is the linear fit to the data yielding a fitted slope of $-$44.9(5) mrad per mm s$^{-1}$, in agreement with the theoretical value of $-$45.3(3) mrad per mm s$^{-1}$ calculated from the average group delay of 6.46(4) $\mu$s over the velocity range using Eq. (2).

Figure 5 (top) shows the phase fluctuations $\Delta \phi$, which determines the velocity sensitivity $\Delta v$=$\Delta \phi$/($k\tau$), of 20 runs with no driving signal to the translation stage. We also estimated the sensitivity with a heated vapor cell at $42.5^\circ$C, and the control field power is chosen close to the maximum group delay of 13.1 $\mu$s at 1.27 mW, as shown in Fig. 5 (bottom). The standard error of the phase uncertainty is 2.9 mrad after 20 runs (32 ms) of integration, which corresponds to a velocity sensitivity of 31 $\mu$m s$^{-1}$ and is equivalent to a short-term sensitivity of 5.5 $\mu$m s$^{-1}$ Hz$^{-1/2}$. The photon shot noise of 20 runs of 70 nW probe power averaging 16 times over a 100 $\mu$s span on the oscilloscope is 9.8$\times$10$^{-6}$ rad, which corresponds to $\Delta v$= 0.11 $\mu$m s$^{-1}$ or 20 nm s$^{-1}$ Hz$^{-1/2}$. Our results are a factor of $\sim$300 times from the photon shot noise limit due to technical noise sources from residual magnetic fields at 50 Hz frequency, air currents on the light path, oscilloscope noise floor, and lossy photodetection, all of which can be improved in the future. Despite being far away from the photon shot noise limit, our results are already 3 orders of magnitude more sensitive than the previous light-dragging experiment, using only a single beam through the vapor cell \cite{Saf} and a cold atoms experiment where $\Delta v$ =1 mm s$^{-1}$ or 6.3 mm s$^{-1}$ Hz$^{-1/2}$ (including the cold atoms preparation time) \cite{Kua}.

\begin{figure}[h]
    \subfigure{\label{fig:5a}\includegraphics[scale=0.3]{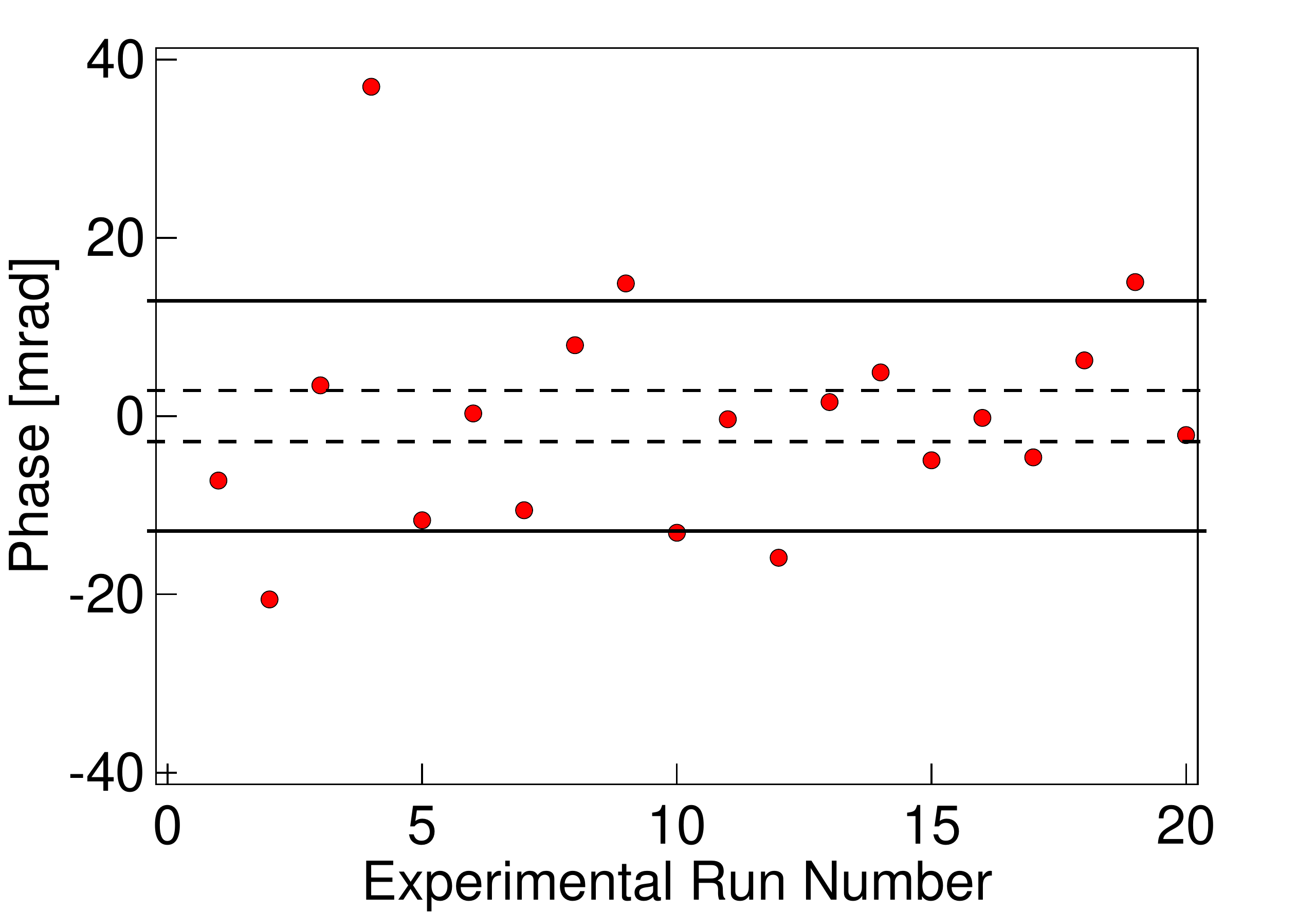}}
    \subfigure{\label{fig:5b}\includegraphics[scale=0.3]{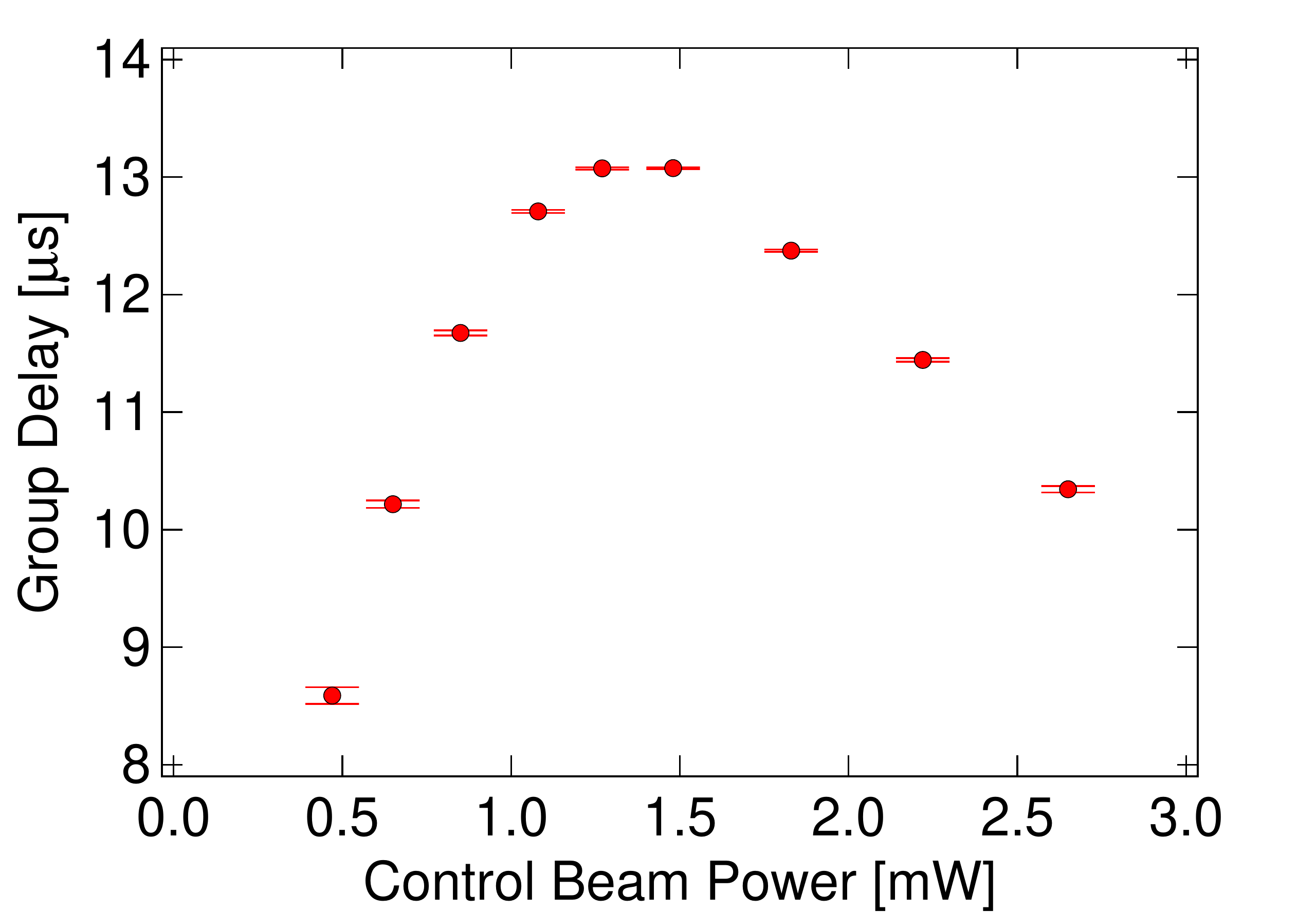}}
  \caption{Phase measurements with heated cell. (Top) Phase uncertainty versus 20 experimental runs. A phase offset has been applied to shift the average to zero. The cell is stationary, and the temperature is maintained at $42.5^\circ$C. The group delay of the probe pulse is 13.1 $\mu$s. The solid lines indicate the standard deviation, and the dashed lines represent the standard error of 2.9 mrad. (Bottom) Group delay of probe pulse in the heated cell versus control beam power. We choose 1.27 mW of control field power for the top figure which is close to the maximum group delay.}
  \label{fig:5}
\end{figure}

We exploit laser-induced quantum interference between two excitation pathways to suppress the absorption of the probe field for motional sensing, which is fundamentally different from the single atoms method in two-photon Raman velocimetry with cold atoms. As the fundamental phase uncertainty is determined by the atom number, the use of thermal atoms collectively allows us to engage $\sim$10$^{10}$ atoms in each shot of phase measurement, compared to $\sim$10$^{6}$ atoms in a typical cold atoms experiment. Our current result improves the resolution of the state-of-the-art two-photon Raman velocimetry of 70 $\mu$m s$^{-1}$ \cite{Cha} by more than a factor of 2, which is equivalent to using a 15 pK cold atomic ensemble if using a single shot Raman pulse. The sensitivity can also be further improved by increasing the group delay of light in the vapor cell. The group delay of a probe field in an EIT medium scales with the ratio of the square root of optical depth to the EIT transmission spectral width. In a paraffin coated vapor cell, 1 kHz linewidth of the transmission spectrum has been observed \cite{Kle}. This may further increase our sensitivity by a factor of 60. In fact, milliseconds of group delay has been achieved by increasing the temperature and length of the vapor cell \cite{Kle2}. The very recent study of atomic spin in hot atomic vapor using $|\Delta m_{\textrm{F}}|$=1 EIT configuration has greatly removed spin-exchange relaxation of atomic coherence and demonstrated storage for 1 s \cite{Kat}. This scheme may improve the stability of our method and reach a level of sensitivity competitive with state-of-the-art atomic motional sensors \cite{Hu}.

We thank Kuan Hong Tan for assistance in the early stage of the experiment. This work is supported by Singapore National Research Foundation under Grant No. NRFF2013-12, Nanyang Technological University under startup grants, and Singapore Ministry of Education under Grant No. MOE2017-T2-2-066.

%\tableofcontents

\nocite{*}

\bibliography{apssamp}% Produces the bibliography via BibTeX.

\end{document}